\documentclass[
reprint,
superscriptaddress,
amsmath,amssymb,
aps,
showpacs
floatfix,
]{revtex4-2}

\usepackage{graphicx}
\usepackage{dcolumn}
\usepackage{bm}
\usepackage[colorlinks,linkcolor=blue,citecolor=blue,urlcolor=blue]{hyperref}
\usepackage[mathlines]{lineno}
\usepackage[colorinlistoftodos]{todonotes}

\begin{document}

\title{X-ray emission associated with radiative recombination for Pb$^{82+}$ ions at threshold energies}

\author{B. Zhu}
\email{binghui.zhu@uni-jena.de or zhubh15@gmail.com}
\affiliation{Helmholtz Institute Jena, D-07743 Jena, Germany}
\affiliation{GSI Helmholtzzentrum für Schwerionenforschung GmbH, D-64291 Darmstadt, Germany}
\affiliation{Institut für Optik und Quantenelektronik, Friedrich-Schiller-Universität Jena, D-07743 Jena, Germany}
\affiliation{School of Nuclear Science and Technology, Lanzhou University, Lanzhou 730000, China}
\author{A. Gumberidze}
\affiliation{GSI Helmholtzzentrum für Schwerionenforschung GmbH, D-64291 Darmstadt, Germany}
\author{T. Over}
\affiliation{Institut für Optik und Quantenelektronik, Friedrich-Schiller-Universität Jena, D-07743 Jena, Germany}\author{G. Weber}
\affiliation{Helmholtz Institute Jena, D-07743 Jena, Germany}
\affiliation{GSI Helmholtzzentrum für Schwerionenforschung GmbH, D-64291 Darmstadt, Germany}
\author{Z. Andelkovic}
\affiliation{GSI Helmholtzzentrum für Schwerionenforschung GmbH, D-64291 Darmstadt, Germany}
\author{A. Bräuning-Demian}
\affiliation{GSI Helmholtzzentrum für Schwerionenforschung GmbH, D-64291 Darmstadt, Germany}
\author{R. Chen}
\affiliation{GSI Helmholtzzentrum für Schwerionenforschung GmbH, D-64291 Darmstadt, Germany}
\affiliation{Institute of Modern Physics, Chinese Academy of Sciences, Lanzhou 730000, China}
\author{D. Dmytriiev}
\affiliation{GSI Helmholtzzentrum für Schwerionenforschung GmbH, D-64291 Darmstadt, Germany}
\author{O. Forstner}
\affiliation{Helmholtz Institute Jena, D-07743 Jena, Germany}
\affiliation{GSI Helmholtzzentrum für Schwerionenforschung GmbH, D-64291 Darmstadt, Germany}
\affiliation{Institut für Optik und Quantenelektronik, Friedrich-Schiller-Universität Jena, D-07743 Jena, Germany}
\author{C. Hahn}
\affiliation{Helmholtz Institute Jena, D-07743 Jena, Germany}
\affiliation{GSI Helmholtzzentrum für Schwerionenforschung GmbH, D-64291 Darmstadt, Germany}
\affiliation{Institut für Optik und Quantenelektronik, Friedrich-Schiller-Universität Jena, D-07743 Jena, Germany}
\author{F. Herfurth}
\affiliation{GSI Helmholtzzentrum für Schwerionenforschung GmbH, D-64291 Darmstadt, Germany}
\author{M. O. Herdrich}
\affiliation{Helmholtz Institute Jena, D-07743 Jena, Germany}
\affiliation{GSI Helmholtzzentrum für Schwerionenforschung GmbH, D-64291 Darmstadt, Germany}
\affiliation{Institut für Optik und Quantenelektronik, Friedrich-Schiller-Universität Jena, D-07743 Jena, Germany}
\author{P. -M. Hillenbrand}
\affiliation{GSI Helmholtzzentrum für Schwerionenforschung GmbH, D-64291 Darmstadt, Germany}
\affiliation{I. Physikalisches Institut, Justus-Liebig-Universität Gießen, D-35392 Giessen, Germany}
\author{A. Kalinin}
\affiliation{GSI Helmholtzzentrum für Schwerionenforschung GmbH, D-64291 Darmstadt, Germany}
\author{F. M. Kröger}
\affiliation{Helmholtz Institute Jena, D-07743 Jena, Germany}
\affiliation{GSI Helmholtzzentrum für Schwerionenforschung GmbH, D-64291 Darmstadt, Germany}
\affiliation{Institut für Optik und Quantenelektronik, Friedrich-Schiller-Universität Jena, D-07743 Jena, Germany}
\author{M. Lestinsky}
\affiliation{GSI Helmholtzzentrum für Schwerionenforschung GmbH, D-64291 Darmstadt, Germany}
\author{Y. A. Litvinov}
\affiliation{GSI Helmholtzzentrum für Schwerionenforschung GmbH, D-64291 Darmstadt, Germany}
\author{E. B. Menz}
\affiliation{Helmholtz Institute Jena, D-07743 Jena, Germany}
\affiliation{GSI Helmholtzzentrum für Schwerionenforschung GmbH, D-64291 Darmstadt, Germany}
\affiliation{Institut für Optik und Quantenelektronik, Friedrich-Schiller-Universität Jena, D-07743 Jena, Germany}
\author{W. Middents}
\affiliation{Helmholtz Institute Jena, D-07743 Jena, Germany}
\affiliation{GSI Helmholtzzentrum für Schwerionenforschung GmbH, D-64291 Darmstadt, Germany}
\affiliation{Institut für Optik und Quantenelektronik, Friedrich-Schiller-Universität Jena, D-07743 Jena, Germany}
\author{T. Morgenroth}
\affiliation{Helmholtz Institute Jena, D-07743 Jena, Germany}
\affiliation{GSI Helmholtzzentrum für Schwerionenforschung GmbH, D-64291 Darmstadt, Germany}
\affiliation{Institut für Optik und Quantenelektronik, Friedrich-Schiller-Universität Jena, D-07743 Jena, Germany}
\author{N. Petridis}
\affiliation{GSI Helmholtzzentrum für Schwerionenforschung GmbH, D-64291 Darmstadt, Germany}
\author{Ph. Pfäfflein}
\affiliation{Helmholtz Institute Jena, D-07743 Jena, Germany}
\affiliation{GSI Helmholtzzentrum für Schwerionenforschung GmbH, D-64291 Darmstadt, Germany}
\affiliation{Institut für Optik und Quantenelektronik, Friedrich-Schiller-Universität Jena, D-07743 Jena, Germany}
\author{M. S. Sanjari}
\affiliation{GSI Helmholtzzentrum für Schwerionenforschung GmbH, D-64291 Darmstadt, Germany}
\affiliation{Aachen University of Applied Sciences, D-52005 Aachen, Germany}
\author{R. S. Sidhu}
\affiliation{GSI Helmholtzzentrum für Schwerionenforschung GmbH, D-64291 Darmstadt, Germany}
\author{U. Spillmann}
\affiliation{GSI Helmholtzzentrum für Schwerionenforschung GmbH, D-64291 Darmstadt, Germany}
\author{R. Schuch}
\affiliation{Physics Department, Stockholm University, S-106 91 Stockholm, Sweden}
\author{S. Schippers}
\affiliation{I. Physikalisches Institut, Justus-Liebig-Universität Gießen, D-35392 Giessen, Germany}
\author{S. Trotsenko}
\affiliation{Helmholtz Institute Jena, D-07743 Jena, Germany}
\affiliation{GSI Helmholtzzentrum für Schwerionenforschung GmbH, D-64291 Darmstadt, Germany}
\author{L. Varga}
\affiliation{GSI Helmholtzzentrum für Schwerionenforschung GmbH, D-64291 Darmstadt, Germany}
\author{G. Vorobyev}
\affiliation{GSI Helmholtzzentrum für Schwerionenforschung GmbH, D-64291 Darmstadt, Germany} 
\author{Th. Stöhlker}
\email{t.stoehlker@gsi.de}
\affiliation{Helmholtz Institute Jena, D-07743 Jena, Germany}
\affiliation{GSI Helmholtzzentrum für Schwerionenforschung GmbH, D-64291 Darmstadt, Germany}
\affiliation{Institut für Optik und Quantenelektronik, Friedrich-Schiller-Universität Jena, D-07743 Jena, Germany}

\date{\today}

\begin{abstract} 

For bare lead ions, decelerated to the low beam energy of 10 MeV/u, the x-ray emission associated 
with radiative recombination (RR) at “cold collision" conditions has been studied at the electron 
cooler of CRYRING@ESR at GSI-Darmstadt. Utilizing dedicated x-ray detection chambers installed 
at 0° and 180° observation geometry, we observed for the very first time for stored ions the full x-ray emission spectrum associated with RR under electron cooling conditions. Most remarkably, no line distortion effects due to delayed emission are present in the well resolved spectra, spanning over a wide range of x-ray energies (from about 5 to 100 keV) which enable to identify fine-structure resolved Lyman, Balmer as well as Paschen x-ray lines along with the RR transitions into the K-, L and M-shell of the ions. To compare with theory, an elaborate theoretical model has been applied. By considering the relativistic atomic structure of Pb$^{81+}$, this model is based on a sophisticated computation of the initial population distribution via RR for all atomic levels up to Rydberg states with principal quantum number $n=$  165 in combination with cascade calculations based on time-dependent rate equations. Within the statistical accuracy, the experimental x-ray line emission is in very good agreement with the results of the theoretical model applied. Most notably, this comparison sheds light on the contribution of prompt and delayed X-ray emission (up to 70 ns) to the observed X-ray spectra, originating in particular from Yrast transitions into inner shells.

\end{abstract}
\maketitle

\section{\label{sec:level1}Introduction}

Radiative recombination (RR), the time-reversal of photoionization,  can be attributed as one of the most elementary and fundamental atomic processes and is the major recombination process occurring for electron beams and bare ions. In particular, it is of utmost importance for matter in highly ionized plasma states, such as those prevalent in stars. In this process, a free electron recombines into a bound state of an ion via the emission of a photon, carrying away the difference in energy  between the initial (continuum) and final (bound) electronic states and satisfying momentum conservation. 

For low relative energies, first measurements of the RR process were done in merged-beams geometry utilizing ion and electron beams \cite{andersen1990radiative,andersen1990rr}. Further detailed experimental studies of this process for low- to mid-Z ions in various electronic configurations  \cite{moller1993aarhus,wolf1991experiments,gwinner2000influence,quinteros1995,gao1995observation} and even many-electron ions at high-Z \cite{lindroth2001qed,schuch2005dielectronic} have been facilitated, in particular, by the advancement of ion storage rings employing electron cooler devices (such as ASTRID, CRYRING, and TSR). For the heaviest ion species these studies were extended at the electron cooler of the ESR ion storage ring \cite{hoffknecht2000recombination,shi2001recombination,reuschl2008state,banas2015subshell}. Moreover, at high relative collision energies, complementary studies have been performed at Electron Beam Ion Trap devices (EBIT) \cite{marrs1994production,marrs1995super}. For completeness we would like to add that very extensive and detailed information about the closely related process of Radiative Electron Capture (REC) occurring in ion-atom collisions has been obtained for a broad range of ion species and collision energies up to ultra-relativistic collision conditions \cite{eichler2007PhysicsReports}. 

Theoretically, RR process has been extensively investigated, first in a semi-classical approach \cite{kramers1923xciii} and then more rigorously by quantum mechanics \cite{stobbe1930quantenmechanik}, and can be treated nowadays in a fully relativistic fashion for initially bare ions \cite{ ichihara1994radiative,eichler1998alignment,surzhykov2003polarization}, whereas for many-electron systems very accurate many-body approaches are also available \cite{scofield1991angular,fritzsche2005radiative}. Moreover, recently, even quantum electrodynamics (QED) contributions have been taken into account in the description of the RR process \cite{shabaev2000qed,holmberg2015qed}. Here, we would like to stress that even though RR has been studied quite intensively over decades and many interesting insights have been gained into various aspects of this process, there are still certain experimental observations which are not yet fully understood \cite{muller1991recombination,gwinner2000influence,gao1995observation,hoffknecht2000recombination,shi2001recombination,banas2015subshell}.

In most of the RR experiments (performed at electron coolers of ion storage rings), total recombination-rates have been measured as a function of the relative energy between electrons and ions. In addition to those, there have been measurements performed at the ESR where x-ray emission due to the RR process in the electron cooler has been observed \cite{beyer1994x, liesen1994x,beyer1995measurement,reuschl2008state,banas2015subshell}. These measurements have provided detailed insights into state-selective population dynamics of the RR and the subsequent cascade processes. Moreover, x-ray spectroscopy at the ESR electron cooler has proven to be a powerful tool to precisely measure binding energies in the heaviest hydrogen- and helium-like ions and thus gain access to QED effects in the regime of strong fields \cite{beyer1994x,beyer1995measurement,gumberidze2004electron,gumberidze2005quantum}. Here, we would like to note that in heavy highly-charged ions the parameter $ v/c \sim \alpha Z $ ($\alpha$ is the fine-structure constant, $v$ being the electron velocity on the ground state) is comparable to unity. Therefore, their basic atomic characteristics, such as energy levels, transition probabilities, are strongly influenced by relativistic or even QED effects. This often leads to new spectral patterns that are remarkably different from the spectra of light elements \cite{mokler1994structure}. To be more specific, the forbidden radiative transitions (such as M1, M2, E2) are enhanced by 2-6 orders of magnitude \cite{marrus1979forbidden,jitrik2004salient} relative to the allowed E1 transition, which scales approximately as $ Z^{4} $.

Recently, CRYRING has been transferred from the Manne Siegbahn Laboratory in Stockholm to GSI/FAIR in Darmstadt and installed behind the ESR storage ring as an in-kind contribution from Sweden to the upcoming international Facility for Anti-proton and Ion Research (FAIR) (see \cite{durante2019all} and references therein). At Stockholm, CRYRING has provided pioneering results over many years \cite{quinteros1995,gao1995observation,lindroth2001qed,schuch2005dielectronic} and has recently been further optimized for future experiments with heavy bare and few-electron ions, exotic nuclei as well as for anti-protons at FAIR \cite{danared2011}. At FAIR, CRYRING@ESR is an important part of a portfolio for trapping and storage facilities for cooled highly-charged heavy ions bridging an energy range over more than 10 orders of magnitude, spanning from rest in the laboratory up to highly relativistic energies \cite{stohlker2015appa}. In combination with the  ESR storage ring, CRYRING offers unique opportunities for a broad range of experiments with electron cooled heaviest one- and few-electron ions at low energies \cite{lestinsky2016physics}. 
At CRYRING@ESR one can store ions at low beam energies (typically below 15 MeV/u), far below the production energy required to produce the desired charge-state of interest e.g. the efficient production of  Pb$^{82+}$ requires at least energies of 300 MeV/u. Another unique selling point of CRYRING@ESR is the adiabatic expansion of the electron beam \cite{danared1994electron,danared2000electron},
resulting in an up to 100 times lower transverse electron temperature as compared to the ESR. Since 2020, the "so called" CRYRING@ESR facility has been fully commissioned both with beams from the local ion-injector as well as from the ESR storage ring and is now available for first physics production runs. 

In this paper, we report on a very first x-ray spectroscopy experiment performed at the electron cooler of the CRYRING@ESR storage ring whereby exploiting dedicated x-ray detection chambers which enabled us to observe x-ray spectra associated with RR of bare lead ions with the cooler electrons in coincidence with down-charged ions at the observation angles of 0° and 180°.  Most importantly, due to the exact 0° and 180° observation geometry, even at 0° x-ray line emission is observed without any line distortion effects caused  by  delayed x-ray emission affecting in particular the Lyman emission  \cite{beyer1994x,liesen1994x,beyer1995measurement,gumberidze2005quantum}.
Moreover, for the new experiment we use beryllium foils for vacuum separation from the 10$^{-11}$ mbar UHV environment of CRYRING@ESR to the x-ray detectors, which enabled us for the first time at an electron cooler to observe the full, well resolved emission pattern ranging from the Paschen series up to the K-RR line under almost background free conditions. These very clean x-ray spectra allow for a detailed comparison of the observed full photon emission pattern with theoretical x-ray spectrum modelling based on elaborate calculations of the RR process, taking into consideration subsequent cascade transitions following RR into excited states of the ion. These detailed studies pave the way for a refined understanding of the level population under such exotic collision conditions, a prerequisite for future precision x-ray spectroscopy studies at the electron cooler of CRYRING@ESR. Such experiments are currently planned within the SPARC collaboration \cite{stohlker2014sparc} aiming to test atomic-structure theory and QED effects in strong fields close to the Schwinger limit with an unprecedented precision by utilizing e.g. novel high-resolution microcalorimeters \cite{pies2012maxs,hengstler2015towards}.

\section{\label{sec:level2}The Experiment}
For the experiment performed in May 2020, about $10^7$ fully stripped Pb ions from the UNILAC/SIS accelerator complex were injected into the ESR at the specific energy of 400 MeV/u. The stored ions were efficiently stochastically cooled at this injection energy as well as by Coulomb interaction with the cold co-moving electrons \cite{poth1990electron} in the ESR electron cooler. Thereafter, the ions were decelerated down to 10 MeV/u, electron cooled again followed by their extraction out of the ESR and injection into the CRYRING@ESR. After these steps, $\sim 10^5$ Pb$ ^{82+}$ ions were usually stored in the CRYRING@ESR. In order to guarantee for a well-defined constant beam velocity, generally of the order of $\Delta\beta/\beta \sim 10^{-5}$, electron currents of typically 10 to 20 mA were applied at the CRYRING@ESR cooler. Moreover, the electron cooling provides a small beam size with a typical diameter of 2 mm, a reduced relative momentum spread $\Delta p/p \sim 10^{-5}$, as well as an emittance of the ion beam of less than $0.1\pi$~mm~mrad.

\begin{figure*}[t]
    \centering
	\includegraphics[scale=0.68]{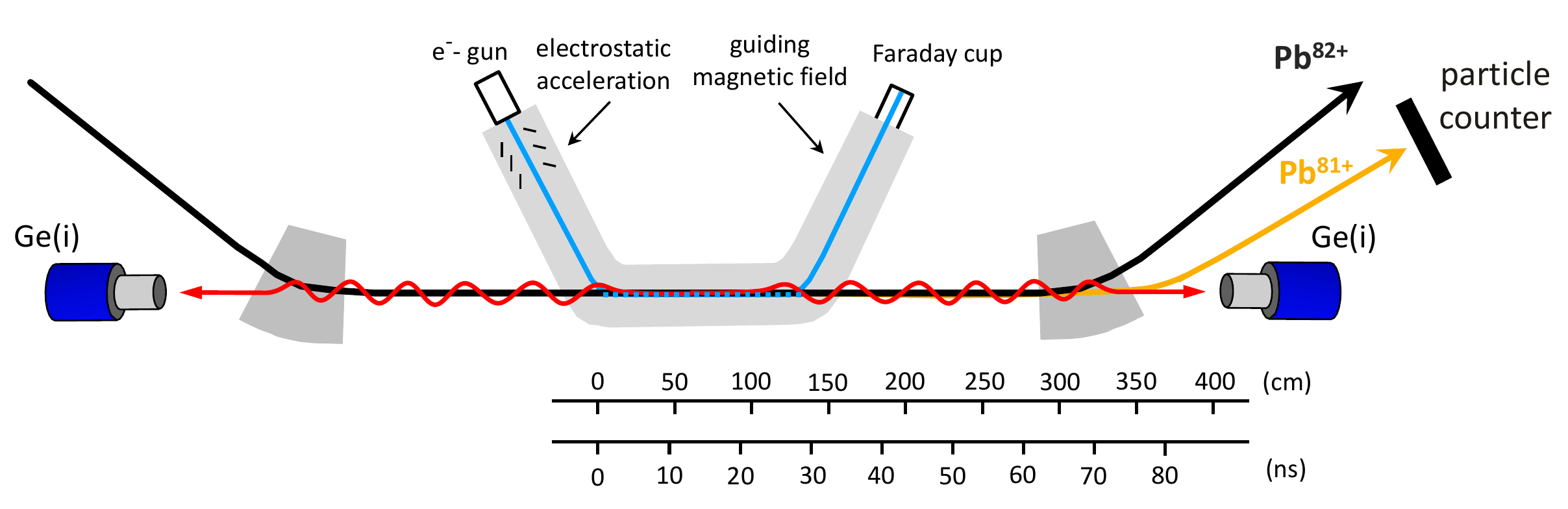}
	\caption{\label{Fig_1} (Color online) Experimental setup at the electron cooler of CRYRING@ESR storage ring. The x-ray emission from the electron-ion beam interaction region is viewed at 0° and 180° by Ge(i) detectors. X-rays are measured in coincidence with down-charged Pb$^{81+}$ projectiles detected by the particle counter installed behind the electron cooler. A length scale and a time scale, valid for a beam velocity of $ \beta \approx 0.146 $ (10 MeV/u), are shown at the bottom. Note, the time of flight of the ions passing the cooler section is close to 30 ns whereas the time of flight of ions from the entrance of the cooler section up to the end of the straight section amounts to 70 ns.}
\end{figure*}

The experimental setup for the measurement of x-ray radiation at the electron cooler of the CRYRING@ESR is shown in Fig. \ref{Fig_1}. At the electron cooler dedicated vacuum separation chambers were used, equipped with beryllium view-ports allowing for x-ray detection under 0° and 180° with respect to the ion beam axis. The exact geometry of the whole detector arrangement was precisely determined by laser assisted trigonometry. The x-ray detection was accomplished by two high-purity, planar germanium x-ray detectors, which were mounted 3.5 m at 0° and 3.3 m at 180° with regard to the midpoint of the roughly 1.2 m long straight electron cooler section. The detector at 0° had a crystal diameter of 16 mm (crystal thickness 10 mm) and was equipped with an x-ray collimator (10 mm thick, made out of brass and reducing the effective detector area to 79 mm$^2$). This results in a solid angle coverage of $\Delta\Omega/\Omega $ which varies  along the cooler section from $ \approx 6.3 \times 10^{-7}$ to  $ \approx 1.3 \times 10^{-6}$, respectively. At  180° the detector used had a diameter of 49.7 mm (crystal thickness 21 mm) and was equipped with an x-ray collimator (10 mm thick, made out of brass and reducing the effective detector area to 962 mm$^2$). There $\Delta\Omega/\Omega $ varies along the cooler section ranging from $ \approx 1.0 \times 10^{-5}$ to $ \approx 4.9 \times 10^{-6}$, respectively. 
In addition, we note that due to the relativistic solid angle transformation, the effective solid angle at 0° observation angle for photons being emitted in the rest frame of the ions is enhanced relative to the one at 180° by a factor of 1.8. In order to suppress the dominant background, stemming from x-ray emission (bremsstrahlung) by a fraction of the electron beam hitting materials in the cooler \cite{schuch1989x} and from natural radioactivity, an ion detector based on secondary electron detection (channel electron multiplier) was operated downstream to the cooler \cite{andelkovic2018}. The coincidences between the x-rays from the interaction region of the electron cooler and the down-charged Pb$ ^{81+} $ ions allowed for the unambiguous identification of those x-rays associated with radiative recombination events.

In the present experiment where recombination of free electrons at zero average relative electron-ion velocity is studied, high Rydberg states of the ions are expected to be populated \cite{beyer1994x, liesen1994x,beyer1995measurement,reuschl2008state,banas2015subshell,pajek1995x}. The ions in these highly excited states might be re-ionized when exposed to magnetic field of the next dipole magnet of CRYRING@ESR on their way to the particle detector. The highest $n_{dip}$ state that survives field ionization, defined by the dipole magnets setting of the storage ring \cite{muller1987experimental}, is usually estimated by the formula \cite{zong1998low,schuch1999recombination}
\begin{equation}
n_{dip}=\left(6.2\times10^{10}q^{3}/v_{i}B\right)^{1/4},
\end{equation}
where $ q $ is the charge state, $ v_{i} $ the ion beam velocity, and $ B $ the magnetic field strength in the unit of T.  In this experiment, the $n_{dip}$ is estimated to be 165 corresponding to the magnetic rigidity value of 1.17 Tm of the dipole magnet. We note that rather sophisticated and detailed studies on the re-ionization of high excited states in magnetic fields have been reported \cite{biedermann1995study,schippers2001storage,bohm2002measurement} pointing out the relevance of low-$l$ levels at high Rydberg states which are not field ionized. Indeed even in our current experiment related cascade transitions could be possible during flight time of ions to the dipole magnet. However, as discussed below these states are not significantly populated via RR (see Fig. \ref{Fig_3} in section \ref{sec:level3}) and will be neglected in the following.

Three-body recombination process in the cooler can lead to population of states with binding energies comparable to the electron beam temperature \cite{beyer1989total,pajek1999plasma} of a few meV. For H-like lead ions, this corresponds to Rydberg state with $n \approx 4000$. Since this value is far beyond $n_{dip}$ and the corresponding radiative lifetime is orders of magnitude larger than the flight time of ions from the center of the cooling section to the dipole magnet, the three-body recombination process will hardly contribute to the observed x-ray intensities \cite{liesen1994x}.

\begin{figure*}[t]
    \centering
	\includegraphics[scale=0.85]{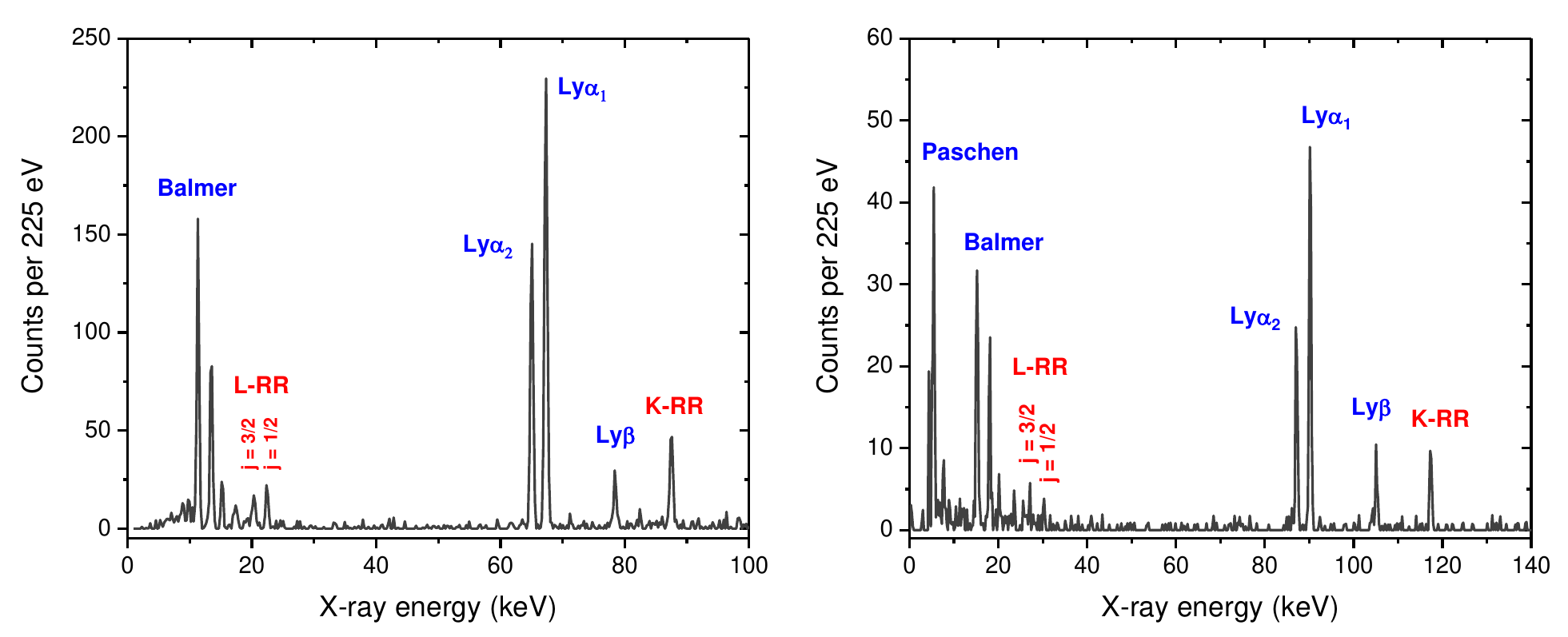}
	\caption{\label{Fig_2} (Color online)  X-ray spectra measured at observation angles of 180° (left side) and 0° (right side) by the two Ge(i) detectors in coincidence with 10 MeV/u  Pb$^{81+}$ projectiles detected after the electron cooler of CRYRING@ESR storage ring. X-ray energies are given in the laboratory frame.}
\end{figure*}

Although in this very first experiment with bare high-Z ions at CRYRING@ESR only a low intensity of $\sim10^{5}$ ions per injection could be achieved, a few days of continuous operation were sufficient to accumulate meaningful coincident x-ray spectra. Here, the experimental study benefits from the low ion beam energy of 10 MeV/u. At this ion beam energy, the energy as well as the intensity of bremsstrahlung caused by the cooler electrons is strongly reduced due to the comparably small cooler voltage of about 5.5~kV and a current of 12~mA  (maximum energy of the bremsstrahlung is given by the electron cooler voltage applied). One may also note that bremstrahlung does not contribute to the coincident (random subtracted) x-ray spectra. Consequently, very clean conditions for x-ray spectroscopy were present at the cooler. Also, our investigation profits from the  0° and 180° geometry of the x-ray detector setup, where Doppler broadening and possible uncertainties of the observation angles are basically not affecting the observed x-ray spectra \cite{gumberidze2005quantum}. The result is depicted in Fig. \ref{Fig_2} where the coincident x-ray spectra are plotted as observed for initially bare lead ions at an energy of 10 MeV/u. The energy scale of the spectra was calibrated during the measurement with  $\gamma$-lines of known energies from standard radioactive calibration sources of $^{241}$Am, $^{133}$Ba, and $^{57}$Co, regularly placed in front of the detectors during the experiment.

\begin{figure*}[t]
    \centering
	\includegraphics[scale=0.43]{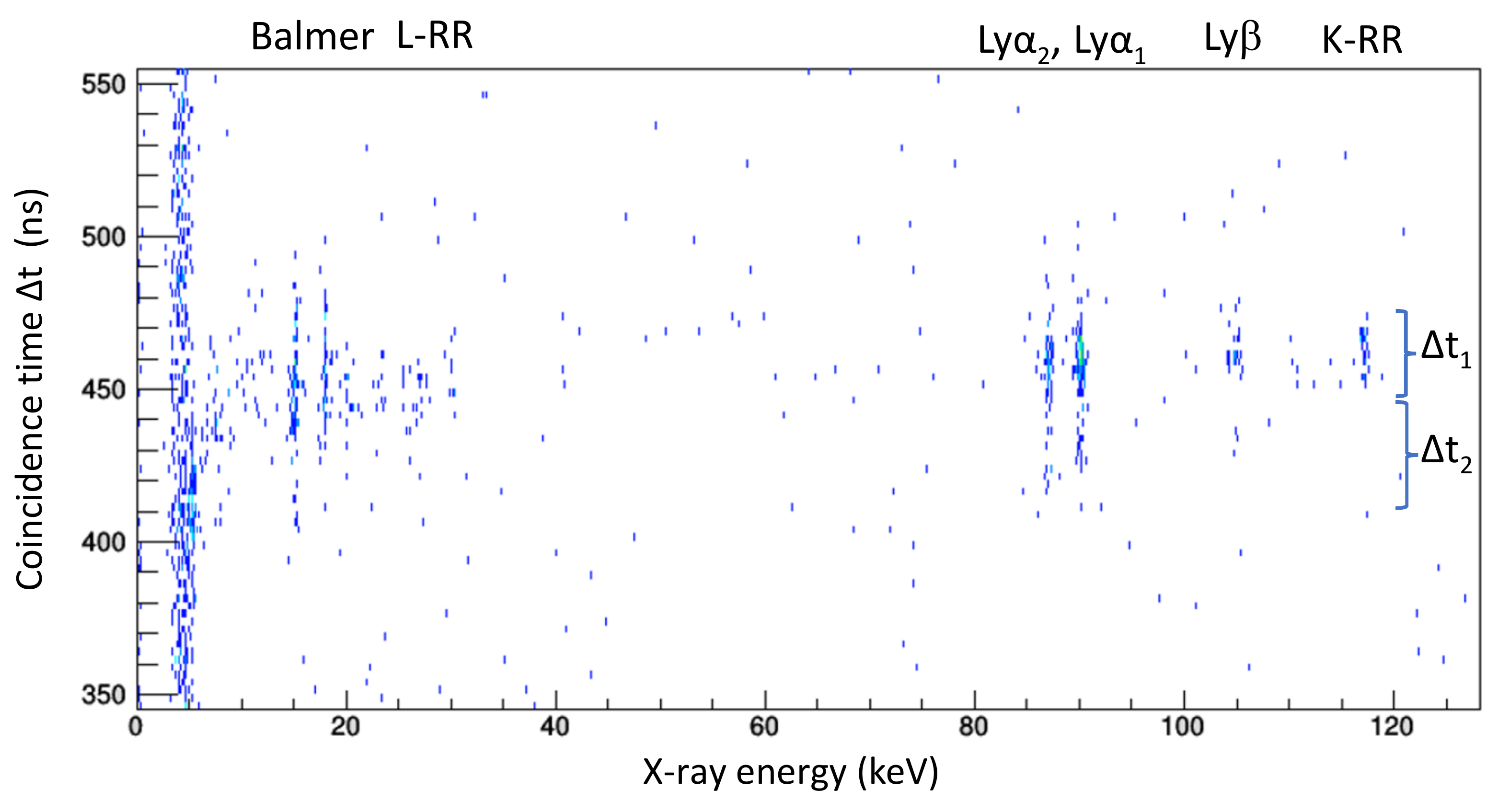}
	\caption{\label{Fig_2a} (Color online)  Two-dimensional presentation of the coincidence time $\Delta t$ versus the x-ray emission observed a 0° (laboratory frame). The coincidence time refers to the time difference between photon (start) and particle (stop) detection (relative time scale). $\Delta t_1 \approx 35$ ns refers to the cooler section; $\Delta t_2 \approx 45$ ns to the region outside the electron cooler.}
\end{figure*}

To shed more light (in a qualitative form) on the interplay between the exact 0° and 180° observation geometry, the extended size of the electron cooler section as well as on the importance of delayed x-ray emission, we display in Fig. \ref{Fig_2a}  a two-dimensional presentation of the coincidence time $\Delta t$ versus the x-ray energy as observed at 0°.  In the figure $\Delta t$ refers to the time difference between photon (start) and particle (stop) detection (relative time scale) and can therefore be interpreted as a time-of-flight spectrum for the x-ray emitting ions (note, the delayed emission is here at earlier times). Moreover, and in contrast to the x-ray spectra shown before, this 2D spectrum refers to all x-ray events registered in coincidence with the particle detector without any  background subtraction applied (which is basically not present for energies above 5 keV).
Therefore, random events caused by bremsstrahlung arising from the cooler section can be identified as a continuous, broad band at an energy close to 5 keV. 
We note that in Fig. \ref{Fig_2a}  we concentrate on purpose on the 0° detector because of its superior timing characteristics which exhibits a time resolution of $\approx 20$~ns for energies above 30 keV. Although the spectrum displayed in Fig. \ref{Fig_2a}  is hampered by statistics, clear line structures are visible corresponding to the line identification in Fig. \ref{Fig_2}  (see also the labels in Fig. \ref{Fig_2a}). Most remarkably, the K-RR radiation, which can only arise from the cooler section of CRYRING@ESR, is distributed over a time interval of about $\Delta t_1 \approx 35$~ns whereas the characteristic radiation (e.g. the Ly$\alpha$ emission) reaches out over a time interval of $\Delta t_1 + \Delta t_2 \approx 80$~ns ($\Delta t_1$ refers to the cooler section; $\Delta t_2$ refers to the straight line outside the electron cooler), enabling a clear identification of x-ray emission occurring outside of the electron cooler region. Considering the estimated flight times for ions inside and outside of the cooler section (see Fig. \ref{Fig_1}), a very good agreement with the experimental findings can be stated. We would like to emphasize that our considerations refer to x-ray energies above 30 keV. Below 30 keV the time response of the x-ray detector deteriorates due to the reduction of the signal to noise ratio. However, extended line structures for the characteristic radiation are clearly visible for the Balmer region too. 

In this context, it is interesting to compare with the situation at the ESR storage ring. In contrast to CRYRING@ESR with its exact 0° and 180° observation geometry, at the ESR electron cooler the x-ray detectors view the cooler section via thin stainless steel foils at an observation angle of close to 0.5° with respect to the ion beam axis \cite{reuschl2008state,banas2015subshell}. As a result, distinctive tails of Lyman-$\alpha$ transition lines are observed at the ESR but not at CRYING@ESR. Note,  these tails are caused by delayed cascade feeding of the L-shell levels as it is also observed at CRYRING@ESR (see Fig. \ref{Fig_2a}). However at the ESR, the delayed  Lyman-$\alpha$ emission occurs at different angles with respect to the x-ray detector (since the detector is not exactly on the beam axis)  \cite{gumberidze2005quantum,reuschl2008state}. 

Finally, we note that the dedicated vacuum chambers mounted for x-ray detection at the electron cooler of CRYRING@ESR were equipped with beryllium windows, improving considerably the transmission for low energy x-rays as compared to the ESR. This enabled the observation of the Balmer and even for the first time the Paschen series, located at the low-energy part of the spectra below 25 keV.

\section{\label{sec:level3}Theoretical background}
In the experiment free electrons are captured via RR into the bare lead ions, populating excited levels of H-like lead which subsequently leads to photon emission by means of deexcitation transitions. Thus, the interpretation of the corresponding x-ray spectra requires an accurate modelling of the RR process at electron cooling conditions based on the integration over the relevant transverse and longitudinal velocity distributions between the ions and the electrons and at the same time to perform appropriate calculations of RR into all relevant levels of the ions whereby considering accurate binding energies and transition rates. 

\subsection{\label{sec:level4}Relevant cross sections and rate coefficients}

The theoretical description of RR is based on considering it as the time reversal of the atomic photoelectric effect. For high-Z ions, such as H-like lead, it has been shown that for low-lying bound states, fully relativistic calculations are necessary to describe the RR process \cite{eichler2007PhysicsReports,brinzanescu1999comparison}. Therefore, in this work, we use complete relativistic calculations for RR into quantum states with $n\leq10$. The exact evaluation of the relativistic photoelectric cross sections requires a partial-wave expansion of the Coulomb-Dirac continuum function. This means that closed-form expressions can no longer be derived, and one has to resort to numerical methods \cite{eichler2007PhysicsReports}. The numerical evaluation of the RR cross sections into strongly bound projectile states were performed in a fully relativistic manner following the detailed formulations in Ref. \cite{pratt1973atomic,eichler1998alignment,ichihara2000cross}. The accuracy of the calculation is mainly determined by the number of expanded partial waves, $\nu=2\kappa_{max}$, with $\kappa_{max}$ being the maximum quantum number for the Dirac angular momentum. For the expansion, a too small choice of $\kappa_{max}$ may lead to truncation errors, whereas a too large value, on the other hand, may lead to explosion of numerical errors due to rapid oscillations of the radial part of the continuum wave function. In our calculations, all multipoles of the electron-photon interaction under inclusion of retardation effects are taken into account.

For the higher excited states with  main quantum number $n>10$ we use the non-relativistic dipole-approximation \cite{stobbe1930quantenmechanik,bethe1957quantum}. In particular, we made use of a set of recurrence relations proposed by Burgess \cite{burgess1960general,burgess1965tables} (for a detailed discussion see \cite{eichler2007PhysicsReports}), allowing one to successively calculate all dipole matrix elements for a given electron kinetic energy, which has the advantage of being numerically stable and applicable for arbitrary $(n,l)$ shells \cite{brinzanescu2001radiative}.

The electrons in the beam have a non-isotropic velocity distribution $f(\mathbf{v})$ in the rest frame of the ions in the cooler setup of a storage ring. In the experiment at the cooler one cannot measure the cross section $\sigma_{nl}$ directly but measures the so-called rate coefficient $\alpha_{nl}$ which is a product of $\sigma_{nl}$ and the relative electron-ion velocity folded with $f(\mathbf{v})$: 
\begin{equation}
   \alpha_{nl}=\left<v\sigma_{nl}(\mathbf{v})\right> =\int v\sigma_{nl}(\mathbf{v})f(\mathbf{v})d^{3}v.
\end{equation}
The measured $\alpha_{nl}$ can then be compared with the one from theoretically calculated $\sigma_{nl}$ and a modelled $f(\mathbf{v})$, usually described in terms of an anisotropic Maxwell-Boltzmann distribution \cite{pajek1992radiative,danared1994electron}, characterized by effective longitudinal $kT_{\parallel}$ and transverse $kT_{\perp}$ beam temperatures (usually in eV):
\begin{equation}
    f(\mathbf{v})=\left(\frac{m_{e}}{2\pi}\right)^{3/2}\frac{1}{kT_{\perp}(kT_{\parallel})^{1/2}}\text{exp}\left[-\left(\frac{m_{e}v_{\perp}^{2}}{2kT_{\perp}}+\frac{m_{e}v_{\parallel}^{2}}{2kT_{\parallel}}\right)\right],
\end{equation}
where $k$ is the Boltzmann constant, and $m_{e}$ the electron mass.

\begin{figure}[h]
    \centering
	\includegraphics[scale=0.75]{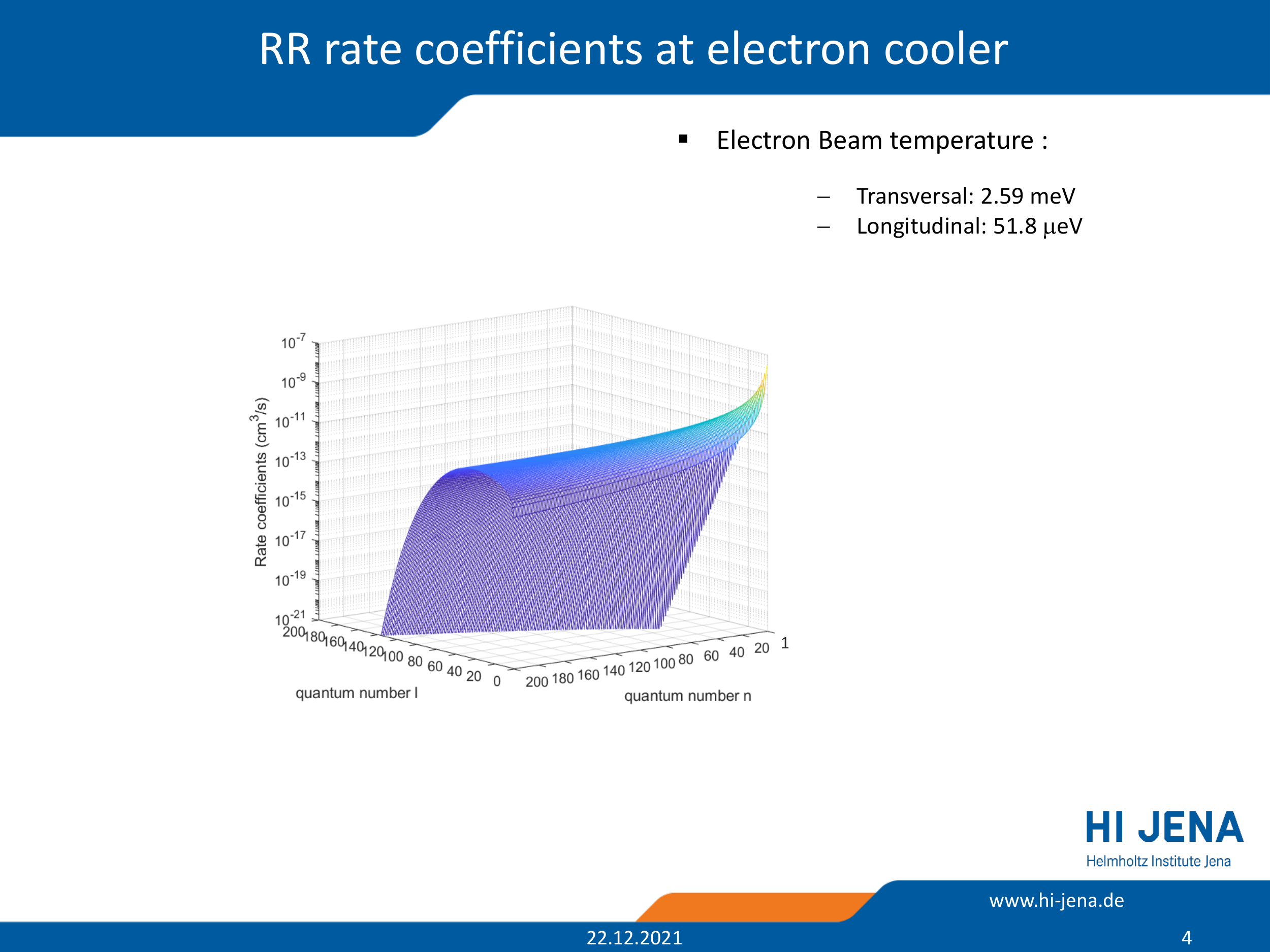}
	\caption{\label{Fig_3} (Color online) State-selective rate coefficients $\alpha_{nl}$ calculated for RR into bare lead ions for flattened electron velocity distribution with $ kT_{\perp} = 2.59 $ meV and $ kT_{\parallel} = 51.8 $ µeV at the CRYRING@ESR electron cooler. A relativistic description of the RR process for inner shells up to $n=10$ is considered here whereas for higher shells the non-relativistic dipole-approximation is applied.}
\end{figure}

Throughout, in the following, an electron beam temperature of about $kT_{\perp} \approx 2.59 $ meV and $kT_{\parallel} \approx 51.8 $ µeV is used for the CRYRING@ESR cooler. These values have been derived from a previous dielectronic-recombination experiment \cite{EsterMenz2021} whereby the overall uncertainties can be estimated to be of the order of 10\%. Figure \ref{Fig_3} depicts the calculated state-selective total rate coefficients for RR into bare lead projectiles at the electron cooler of CRYRING@ESR as a function of the principal and the angular momentum quantum number $\left(n,l\right)$. As can be seen, due to the small relative velocity between the electrons and the ions, the RR rate coefficient decreases only slightly with increasing principle quantum number $n$ ($\alpha_{nl} \sim 1/n$) \cite{gao1996role}. For completeness we note, that for high Rydberg states with the binding energies $E_B <  kT_{\perp}$  the scaling law for RR at high relative energies is restored again ($\alpha_{nl} \sim 1/n^{3}$) \cite{stohlker1997strong}. Correspondingly, there are very significant contributions from high $\left(n,l\right)$ levels regarding the production of characteristic projectile x-rays resulting via radiative cascades from these high Rydberg levels \cite{reuschl2008state,pajek1995x}.

\begin{figure*}[t]
    \centering
	\includegraphics[scale=0.85]{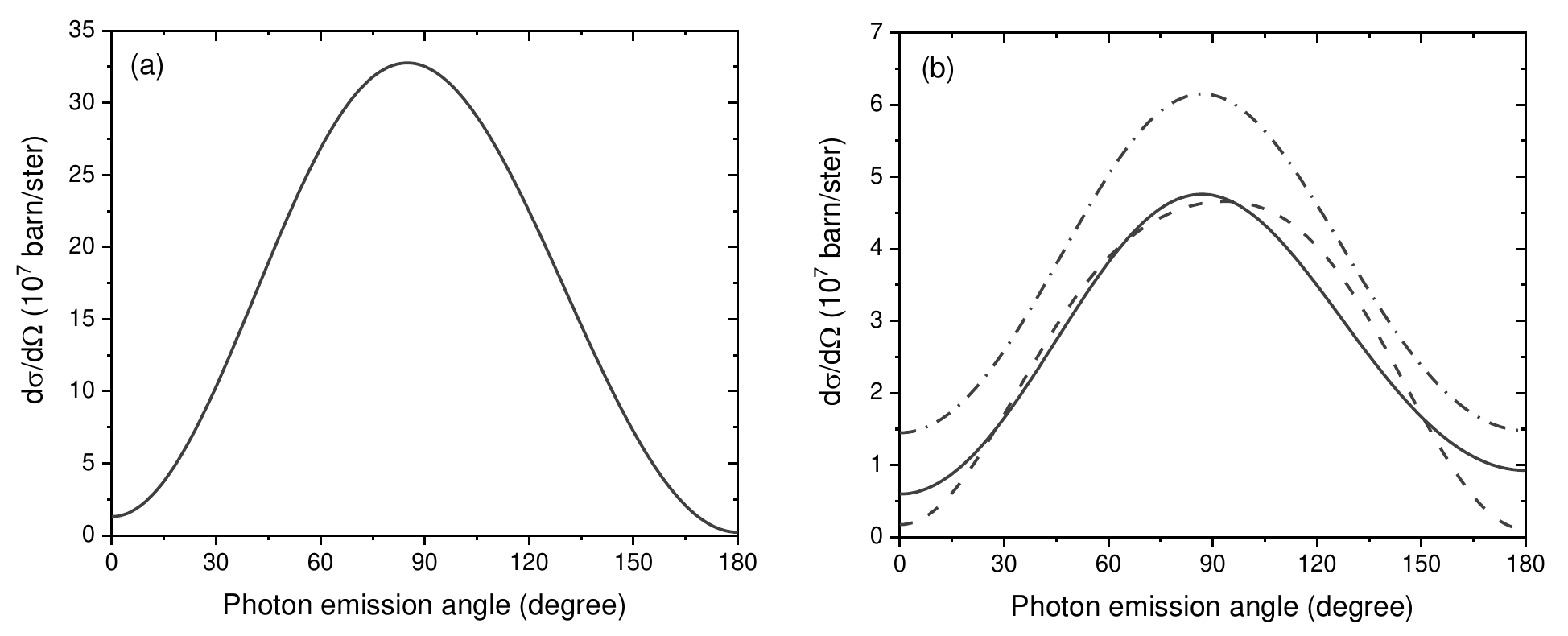}
	\caption{\label{Fig_4} (Color online) Differential cross sections per vacancy calculated within the framework of the rigorous relativistic theory applied for RR of free electrons with bare lead ions at relative electron-ion energy of 2.59 meV. (a) For K-shell and (b) For the various levels of the L-shell (full line: 2p$_{3/2}$; dash-dotted line: 2p$_{1/2}$; dashed line: 2s$_{1/2})$.}
\end{figure*}

In order to describe our experimental data, angular-differential RR rate coefficients are needed since the electron velocity distribution is non-isotropic in the rest frame of the ion and our x-ray detectors are placed at specific observation angles (0° and 180° with respect to the ion beam direction). We performed relativistic calculations of the differential RR cross sections for the inner shells based on the technique as discussed in detail by Eichler et al. \cite{ichihara1994radiative,eichler1995photon,ichihara2001angle,eichler2007PhysicsReports} in order to describe the observed direct RR transitions into the K-, L-, and M-shell of the ions. As an example, we show in Fig. \ref{Fig_4} the angular-differential cross sections for RR into K- and L-shell of initially bare lead ions. In the particular case of the RR process occurring in the electron cooler, the transverse temperature ($kT_{\perp}=2.59 $ meV) of the electron beam is much larger than the longitudinal one ($kT_{\parallel}=51.8 $ µeV), and thus the collision axis is defined as the one perpendicular to the ion beam axis. Correspondingly, the calculated RR rate coefficients at 90° are used to compare with the experimental data registered by the two x-ray detectors mounted at 0° and 180° along the ion beam axis. Here, it is important to stress that the shape of the RR angular distribution remains largely unchanged for the whole low-collision energy range defined by the temperatures of the electron beam in the electron cooler \cite{eichler2007PhysicsReports}. 

As already mentioned, at the collision energies of a few meV as they are prevailing at the electron cooler of CRYRING@ESR, recombination into high Rydberg states and following radiative cascades will contribute significantly to the intensity of the characteristic radiation \cite{beyer1994x,liesen1994x,reuschl2008state} (see also Fig. \ref{Fig_3}). This will substantially attenuate any potential initial alignment of excited states produced via RR which finally populate the states of the L- and M-shell states via radiative cascades. Therefore and in contrast to the prompt RR radiation, we assume an isotropic emission pattern in the projectile frame for the subsequent characteristic projectile transitions of the observed Lyman, Balmer, and Paschen series \cite{eichler1994angluar,stohlker1997strong}.

\subsection{\label{sec:level5}Transition energies and probabilities}
Since no existing database can provide the required extensive set of energy levels and radiative transition rates up to $n=165$ for the relevant ions of lead, the input data for modelling of the observed x-ray spectra need to be generated using an accurate atomic code. The Flexible Atomic Code (FAC) developed by M. F. Gu \cite{gu2003indirect,gu2008flexible} based on the relativistic configuration interaction method with independent-particle base wave-functions was used to generate the required atomic data, in which QED corrections are treated as hydrogenic approximations for self-energy and vacuum polarization effects. For H-like lead ions, the comparison of energy levels obtained from FAC code with the results from Johnson and Soff \cite{johnson1985lamb} as well as from Yerokhin and Shabaev \cite{yerokhin2015lamb} result in an accuracy up to a few eV, and the radiative transition rates are accurate to $99 \ \%$ when compared with tabulated data from Pal’chikov \cite{pal1998relativistic}.

For high Rydberg states with the main quantum number  $n>100$, a fast computation code \cite{storey1991fast} was applied for the evaluation of radiative properties in non-relativistic approximation. For bound-bound transitions, the technique based on recurrence relations calculating the dipole matrix elements had proven to be accurate and stable for values of the main quantum number $n$ of up to 500. 

\section{\label{sec:level6}Results and discussion}
As discussed in the previous section, radiative recombination into highly excited states of the projectile will result in decay cascades, mainly by electric dipole transitions, and is likely to end up in one of the intermediate states considered here. While the cascade photons between highly excited levels are usually not detected, the decay photons involving K-, L-, and M-shells are measured by the Ge(i) detectors (see Fig. \ref{Fig_2}).
The measured x-ray spectra were corrected for the energy dependent detection efficiencies of the individual x-ray detectors. For this purpose the germanium detector response function was simulated with the well-established Monte Carlo EGS5 code \cite{hirayama2005egs5}. Note, for the specific detectors applied this method has been proven to provide reliable results in particular for the case of relative (not absolute) detector efficiencies (see e.g. \cite{banas2013two}). 

In the following, we concentrate on the discussion of the observed prompt RR transitions as well as of the characteristic x-ray lines in comparison with the simulation based on the theoretical modelling described in section \ref{sec:level3}. One may note that the very small line broadening of the RR lines due to the temperature of the electron beam is negligible in comparison with the intrinsic resolution of the detectors. The strongest line associated with the direct RR refers to recombination into the 1s$_{1/2}$ ground state (K-RR) and the intensity of other recombination transitions observed drop off as the recombination rates scale approximately as $1/n$ \cite{kramers1923xciii,gao1996role} for the inner shells. The L-RR populates the excited L-shell levels in Pb$^{81+}$, and hence contributes to the initial line intensity of the  Ly$\alpha_{1,2}$ transition. The M-RR a.s.o contributes to the Balmer series and the Ly$\beta$ transition. 

\begin{figure*}[t]
    \centering
	\includegraphics[scale=0.96]{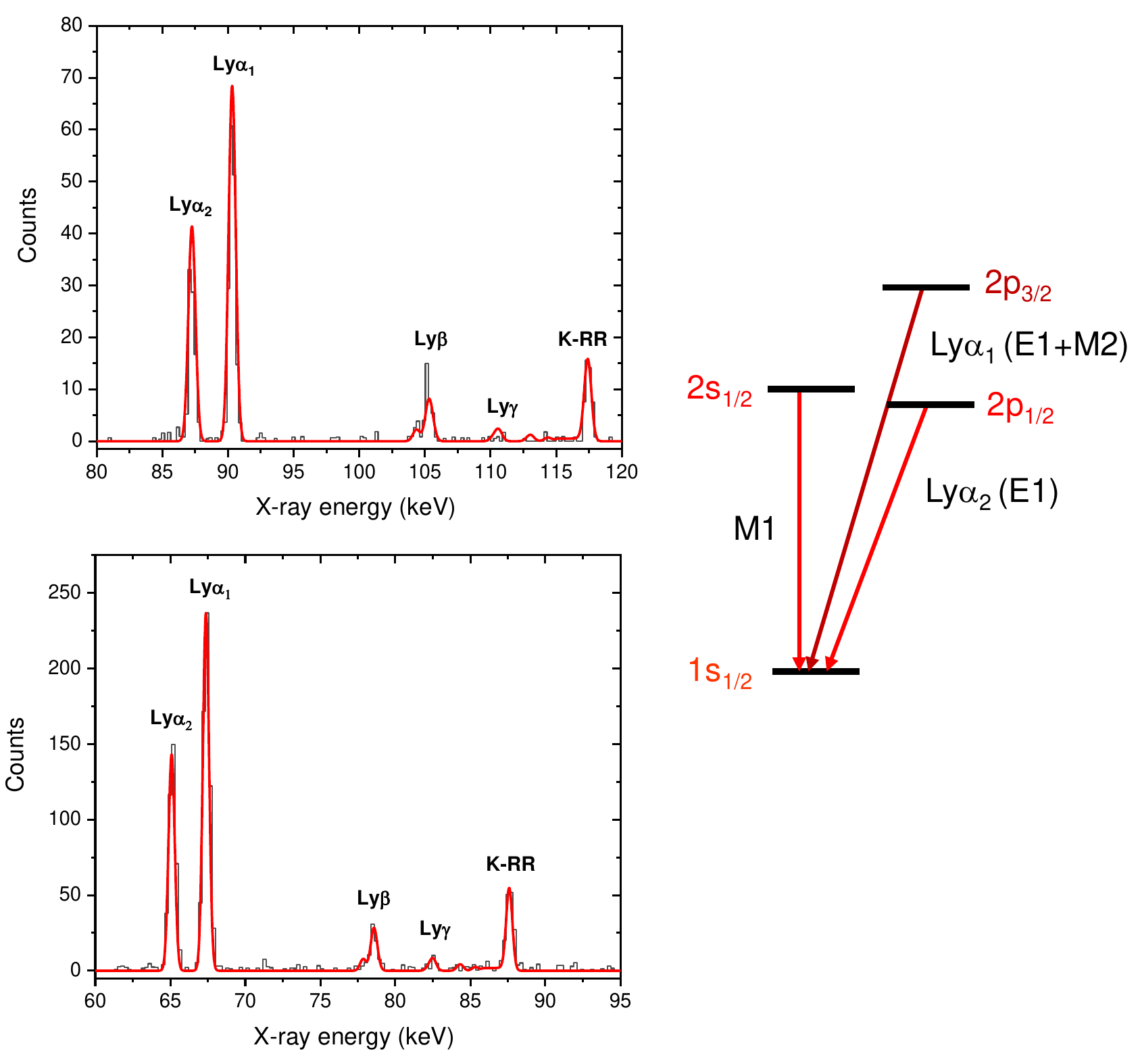}
	\caption{\label{Fig_5} (Color online) The Lyman series and the K-RR measured at 0° (top part) and 180° (bottom part) relative to the ion beam axis (thin grey line). The thick red line displays results of the simulation (for details see text). The intensity of the K-RR line is used for normalization. In the level diagram all possible transition types between the states with $n=1, 2$ in H-like lead are shown. }
\end{figure*}

\begin{figure*}[t]
    \centering
	\includegraphics[scale=0.9]{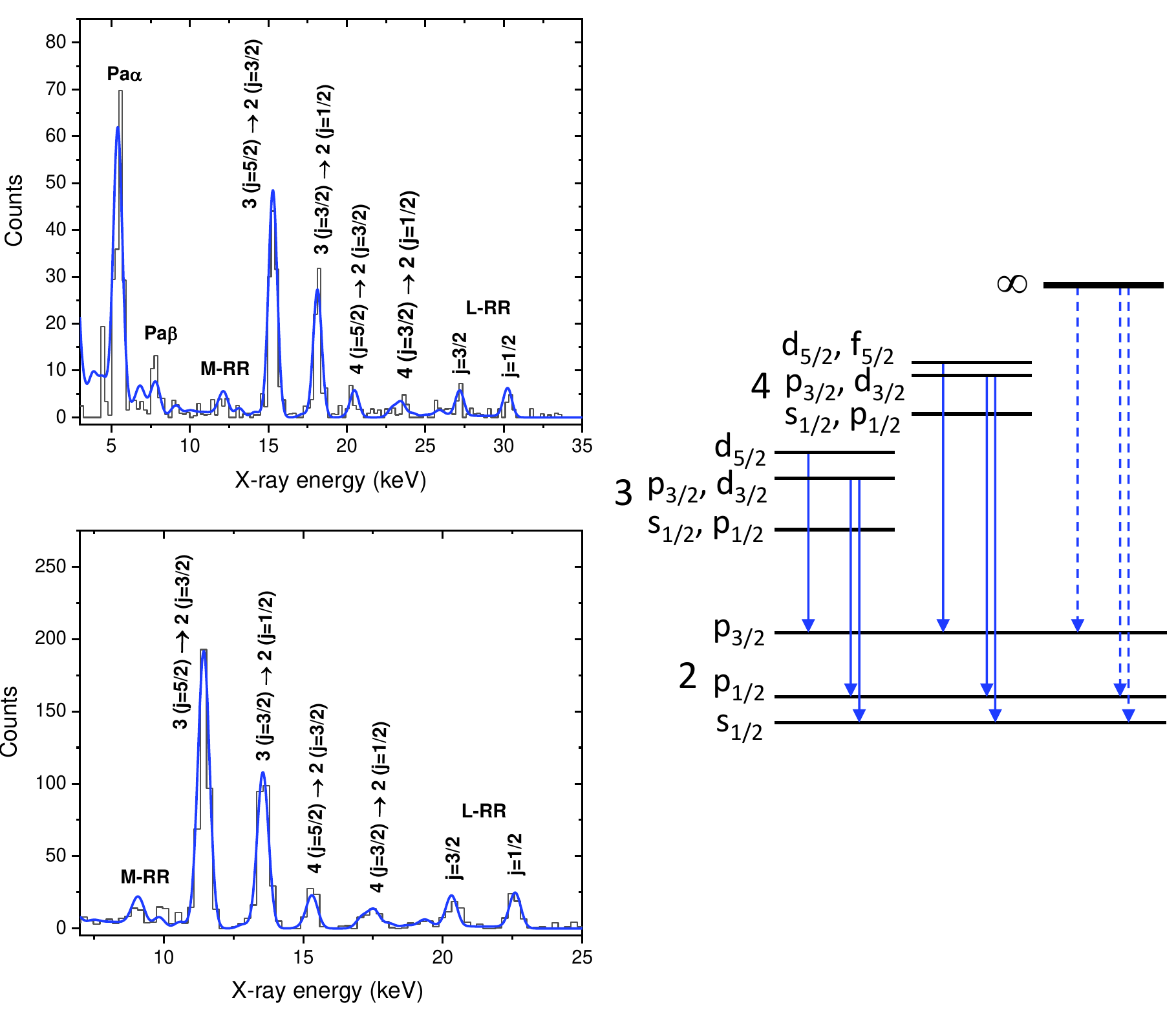}
	\caption{\label{Fig_6} (Color online) X-ray spectra associated with prompt L-RR and M-RR radiation together with characteristic Balmer and Paschen transitions measured by two Ge(i) detectors placed at 0° (top part) and 180° (bottom part) with respect to the ion beam axis. The blue line gives the results of the cascade simulation. Two intense Balmer transitions are Ba$\alpha_{1}$: $3 \ (j=5/2)\rightarrow2 \ (j=3/2) $ and Ba$\alpha_{2}$: $3 \ (j=3/2)\rightarrow2 \ (j=1/2) $.  In the level diagram the Balmer transitions relevant for our measurement are shown.}
\end{figure*}

Figure \ref{Fig_5} shows the part of the x-ray spectra containing the Lyman transitions along with the K-RR radiation (laboratory frame; thin solid line), recorded at the two observation angles of 0° and 180° in comparison with the result of the spectra simulation (thick solid line). The simulated spectral line profiles are approximated by a Gaussian function with a FWHM of 550 eV, to account for the intrinsic resolution of the Ge(i) detectors used. In addition, the experimental spectra displayed have been corrected for the energy-dependent detector (relative) efficiency.
The prompt K-RR line intensity appearing at an x-ray energy of 101 keV in the projectile frame was used to normalize the simulated Lyman spectrum. Note, for the Ly$\alpha_{2}$ 2p$_{1/2} \rightarrow $ 1s$_{1/2}$ transition there is a line blend due to the 2s$_{1/2} \rightarrow $ 1s$_{1/2}$ M1 decay which is energetically separated from the former by the $n=2$ Lamb-shift of 39 eV \cite{yerokhin2015lamb} which cannot be resolved in our spectra due the intrinsic detector resolution. Moreover, we note that for Pb$^{81+}$ the M1 decay rate amounts to $ \Gamma_{\text{M1}}=5.32 \times 10^{13} $ s$^{-1}$ whereas the one for the competing two-photon decay channel to $\Gamma_{\text{2E1}}=2.03 \times 10^{12} $ s$^{-1}$ \cite{parpia1982radiative}. Due to the dominance of the M1 decay for the 2s$_{1/2}$ state, in the following the two-photon decay is neglected. Based on our spectrum simulation, we estimate that the M1 2s$_{1/2} \rightarrow $ 1s$_{1/2}$ ground state transition contributes by about 15$\%$ to the observed Ly$\alpha_{2}$ intensity. Overall, a very good agreement between the experimental and the simulated spectra can be stated for both observation angles.

Complementary information on the Lyman spectrum is provided by the Balmer and Paschen series. In Fig. \ref{Fig_6} we compare the experimentally recorded low-energy x-ray spectra (thin line), consisting of the prompt L-RR and M-RR x-ray transitions together with the characteristic Balmer and Paschen series at 0° and 180°, with our theoretical model based on cascade calculations (thick line). In contrast to the Lyman spectra where K-RR has been used for normalization, here, due to the relatively low statistics in the L-RR lines, the dominant transition line $ \left(3 \ (j=5/2)\rightarrow2 \ (j=3/2)\right) $ appearing at an x-ray energy of 13 keV in the projectile frame (corresponding to 15 keV at 0° and 11 keV at 180°, respectively) was used to normalize the simulated spectra.  
In addition, the most prominent Balmer lines in the spectra are marked by arrows in the level diagram, explaining the origin of various characteristic x-ray lines. It has to be stressed that, due to the Doppler blue-shift in combination with the high transmission of low-energy photons through the beryllium  view port at 0°, Paschen radiation from Pb$^{81+}$ was observed for the first time with a relative intensity comparable to the overall line intensities of the Balmer spectrum.  However, due to the narrow line spacing, the experimental resolution does not allow us to resolve the individual transitions contributing to the observed Paschen-$\alpha$ line whereas in the case of the Balmer series, a multitude of transitions, basically due to the fine structure splitting of the $n=2$ and $n=3$ states, are clearly visible in both x-ray detectors. We note that the also well resolved L-RR$_{j=3/2}$ and L-RR$_{j=1/2}$ x-ray lines mark at the same time the series limit for transitions decaying from high $n$ levels directly into the 2s$_{1/2}$, 2p$_{1/2}$ and 2p$_{3/2}$ levels, respectively. 

In contrast to the Balmer series measured at high collision energies where RR favors capture into s-states resulting in $s\rightarrow p$ transitions \cite{stohlker1997strong} (e.g. the 3s$_{1/2}\rightarrow$ 2p$_{3/2}$), the dominant transitions observed in the current experiment stem from atomic levels with angular momenta $l \geq 1$. 
This distinct difference is again attributed to the role of the RR population mechanism in which electron capture at low relative energies populates preferentially $n$ states with $l \approx n/3$ \cite{pajek1992radiative}. 
 Moreover, for comparing with the observed x-ray line emission, the initial $(n,l)$ population distribution is the starting point for cascade calculations based on the specific decay rates for the individual levels to be considered. For the inner shells ($n <$ 10), these cascades lead to a preferred population of high angular momentum states with $l = n -1$, resulting in subsequent Yrast transitions with $l \rightarrow l - 1$ \cite{reuschl2006balmer}. We note, that the corresponding theoretical x-ray line spectra seem to describe the experimental findings very well (see Figs. \ref{Fig_5} and \ref{Fig_6}). 
\begin{figure}[h]
    \centering
	\includegraphics[scale=0.91]{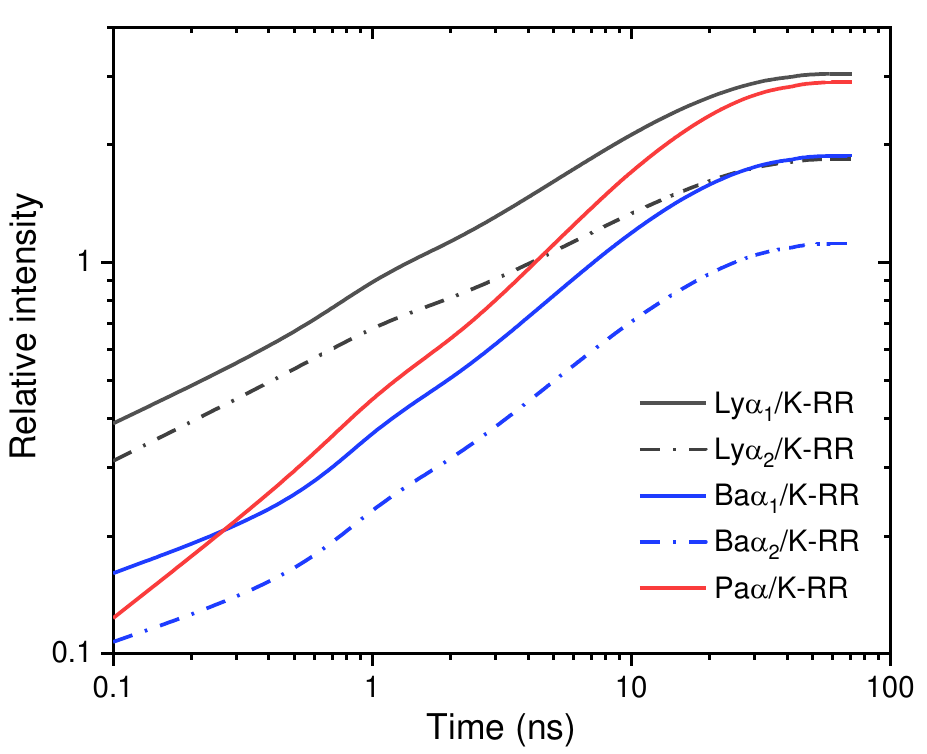}
	\caption{\label{Fig_7} (Color online) Cascade calculations of the characteristic line intensities as a function of flight time of charge exchanged Pb$^{81+}$ ions at 10 MeV/u at the electron cooler of the CRYRING@ESR storage ring. All line intensities are normalized to the K-RR population. Calculations are performed for recombination into excited projectile states up to $n=165$.}
\end{figure}

\begin{figure}[h]
    \centering
	\includegraphics[scale=0.91]{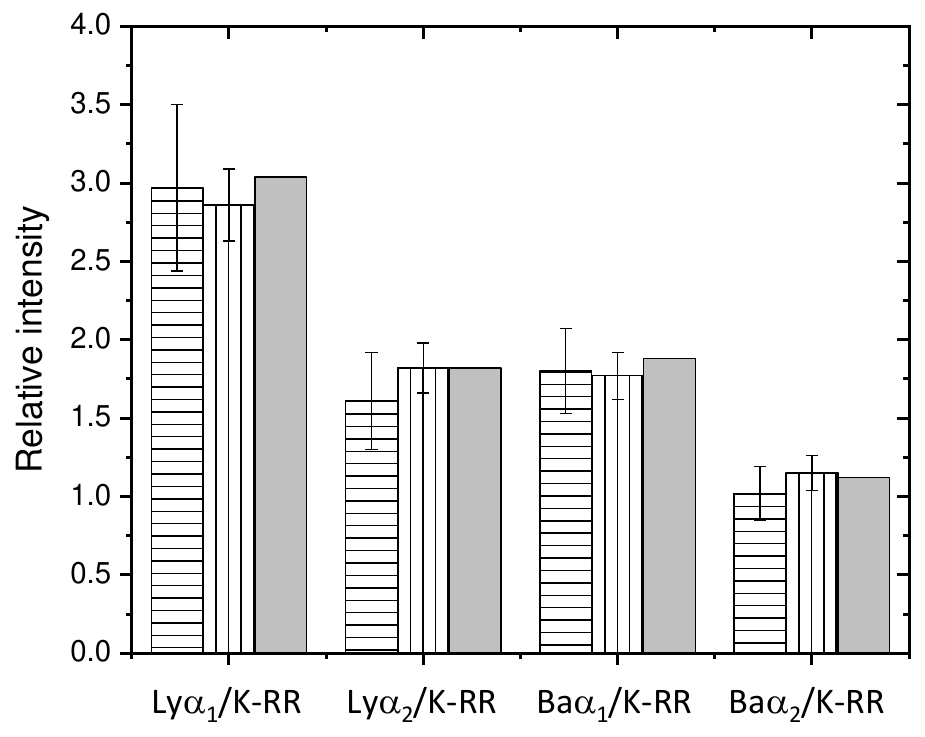}
	\caption{\label{Fig_8} (Color online) Experimental results in comparison with the time-dependent theoretical model (grey columns) for characteristic line intensities normalized to the K-RR population. White columns with horizontal strips show experimental data at 0°, white columns with vertical strips show experimental data at 180°.}
\end{figure}

Since we record in our measurement both the prompt RR transitions as well as characteristic Lyman, Balmer, and even Paschen series produced to a large extent via RR into high Rydberg states and subsequent cascades, our spectra provides a possibility to study the time development of the mentioned cascade process. The calculated time development of the characteristic line emission is depicted in Fig. \ref{Fig_7}. In the figure, the characteristic line intensities are normalized to the prompt K-RR intensity. As can be seen, the initial characteristic line intensities induced by direct RR into L-, and M-shells (t=0) are enhanced by up to a factor of 10 following deexcitation cascades within  $\Delta t < $ 40 ns. In general, we can state that within this time period, to be compared to the time of flight of the ions inside the cooler section of $\approx 30 $ ns, the vast majority of electrons having recombined into excited states of the ion reach the ground state. 
Here, we would like to emphasize that this time is comparable to the time of flight of the ions ($\approx 40 $ ns) from the cooler to the dipole magnet in front of the 0-degree detector (see Fig. \ref{Fig_1}).  Indeed, the time integrated line intensities for both detectors appear within the statistical accuracy to be in excellent agreement with the results from cascade calculations, as shown in Fig. \ref{Fig_8}. 
To underline these findings we present in Table \ref{table1} the experimental ratios of the L-RR fine structure intensities normalized to the one for K-RR in comparison to our theoretical modelling along with the observed Ly$\alpha$ intensities (time integrated) also normalized to the one of K-RR. These intensity ratios should represent the normalized Lyman-$\alpha$ intensities at t=0 (prompt x-ray emission) in comparison to the time integrated ($\Delta t = 70$ ns) one. 
Here, we would like to note that for comparison with theory, in Fig. \ref{Fig_8} and Table \ref{table1}, the experimental intensity ratios have been corrected for the detection efficiency for both x-ray detectors. It is also important to stress that these findings are generally consistent with those from earlier studies conducted at the ESR electron cooler \cite{beyer1994x,liesen1994x,reuschl2008state,pajek1995x}.

\begin{table*}[t]
	\centering
	\caption{The Ly$\alpha_{1}$/K-RR and Ly$\alpha_{2}$/K-RR ratios obtained for RR into Pb$^{82+} $ in comparison with the time-dependent theoretical model. Corrections on detection efficiencies and angular distributions are applied. Uncertainties shown here are due to counting statistics whereas systematic uncertainties are neglected.}
	\setlength{\tabcolsep}{14pt}
    \renewcommand{\arraystretch}{1.1} 
	\scalebox{1.0}{
		\begin{tabular}[t]{ccccc}
			\hline\hline
			 &   & 0 deg & 180 deg & Theory  \\
			\hline
		    $t \sim 0 $ ns & $ j=3/2 $ & 0.287 $\pm$ 0.136 & 0.289 $\pm$ 0.061 & 0.304 \\
		     & $ j=1/2 $ & 0.338 $\pm$ 0.151 & 0.378 $\pm$ 0.072 & 0.361 \\
		    \hline
		    $\Delta t \sim 70 $ ns & $ j=3/2 $ & 2.965 $\pm$ 0.528 & 2.852 $\pm$ 0.229 & 3.042 \\
		     & $ j=1/2 $ & 1.613 $\pm$ 0.316 & 1.821 $\pm$ 0.156 & 1.843 \\		     
			\hline\hline
	\end{tabular}}
	\label{table1}
\end{table*}

Based on the comparison with the observed Lyman-$\alpha$ line intensities (see Fig. \ref{Fig_8}), we conclude that there is a very  significant contribution to the  Lyman-$\alpha$  emission arising  from recombination into highly excited states and subsequent cascades: 86.3$\%$ $\pm$ 15.6 $\%$ for Ly$\alpha_{2}$ and 93.6$\%$ $\pm$ 12$\%$ for Ly$\alpha_{1}$, respectively for the 0° detector; 87.4$\%$ $\pm$ 6.5$\%$ for Ly$\alpha_{2}$ and 94$\%$ $\pm$ 5.5$\%$ for Ly$\alpha_{1}$ at 180°, respectively. Also we would like to point out that within the statistical accuracy reached, the consistent results obtained at 0° and 180° appear to be kind of remarkable because of the very different observation geometries. Whereas the observation at 0° is in particular sensitive to delayed emission occurring outside the electron cooler section, the latter is substantially suppressed at 180° due to the strongly reduced solid angle (see section \ref{sec:level2}). This also confirms the theoretical model applied which predicts that strongly delayed x-ray emission due to recombination into high Rydberg states contributes only very minor to the total recombination rate for the excited states (e.g. recombination into excited states with principal quantum numbers below $n=100$ contributes to 90 \% to the total rate).

\section{\label{sec:level7}Summary and outlook}
In summary, we have performed the first x-ray spectroscopic investigation of the RR process for bare lead ions at the electron cooler of the CRYRING@ESR storage ring which has recently been installed and commissioned at GSI/FAIR. Coupling of CRYRING@ESR with the ESR enables for the first time storage of the heaviest bare ions in CRYRING@ESR and thus studying the x-ray emission associated with RR into heavy bare ions with the ultra-cold electrons of the CRYRING electron cooler. 
This is made possible by deceleration of the Pb$^{82+}$ beam in the ESR from the injection energy of 400 MeV/u to 10 MeV/u, followed by subsequent transfer, storage, and cooling at CRYRING@ESR as well as by the use of dedicated x-ray detection chambers at the straight cooler section of the storage ring at 0° and 180° observation geometry.

The sophisticated, time-dependent modelling of the observed x-ray spectra enables to reproduce qualitatively the detailed spectral information for RR of cooler electrons with the bare lead ions. Beside the prompt recombination transitions for RR into the K-, L-, and M-shells, all characteristic x-ray emission features such as the observed Lyman, Balmer, and Paschen lines are well reproduced by the applied spectrum modelling. We note, that these detailed findings were made possible by the dedicated x-ray detection chambers at 0° and 180° installed at the cooler section equipped with beryllium vacuum windows enabling a high transmission even for x-rays in region close to a few keV, along with the excellent, overall performance of CRYRING@ESR providing very well defined electron cooled Pb$^{82+}$ beams at an energy of as low as 10 MeV/u. Furthermore, the precisely defined observation geometry enabled to observe all x-ray lines without any line distortion effects and basically without any Doppler broadening, the performance of CRYRING@ESR guaranteed for cooled ion beams by applying electron cooler currents of as low as 12 mA avoiding a high-level of a potentially disturbing x-ray background. As a consequence, even the complete Balmer series as well as Paschen lines of a high-Z element could be observed for the very first time in a well resolved fashion  at a cooler section of a storage ring. Based on these findings we conclude that RR at electron cooling can be utilized for detailed precision spectroscopy of heavy highly charged ions.

At such low relative beam collision energies, the relative $(n,l)$ population rate coefficients are only slightly affected by a variation of the cooler temperature since radiative recombination at low relative energies will always favor high quantum states  leading finally to an Yrast cascade as observed in the current experiment. In addition, we note that a large fraction of electrons transferred via RR into excited levels of the ions have reached the ground state within our experimental observation time of about $\approx$ 70 ns. Indeed, our experiment detects the major contribution to the x-ray emission that arises from RR into states with principal quantum numbers up to around $n=100$.

In order to reach a sensitivity to higher Rydberg states, a substantial improvement of the statistical accuracy is required. In addition, one may aim at the measurement of absolute x-ray line intensities in contrast to intensity ratios as applied in the present study. This may enhance dramatically the sensitivity reached  and is finally required to study subtle details of recombination processes in the electron cooler based on x-ray spectroscopy. A further challenging aspect of future studies would be the exploitation of the timing characteristics for recombination into high-Z ions. In the presented study, the time of flight of ions passing the electron cooler, the drift time of the ions from the electron cooler to the dipole magnet located downstream of the cooler as well as typical time interval for the electron cascade are all in the regime between 30 and 40 ns. In future experimental studies one may consider to decelerate the ions even further inside of CRYRING@ESR to energies of 5 MeV/u or even below. This would drastically prolong the time of flight of the ions and enable time of flight with high sensitivity (between the x-rays and the particle detector for the down-charged ions), allowing us to separate experimentally delayed and prompt x-ray emission in a clear cut fashion. 

In conclusion, the present study demonstrates the favorable experimental conditions for x-ray spectroscopy at the electron cooler of CRYRING@ESR 
for precision atomic-structure studies for high-Z one- and few-electron ions and may enable in the near future the exploitation of high-resolution crystal spectrometers or even highly-granular, novel high-resolution microcalorimeters. 

\begin{acknowledgements}
The substantial support provided by Norbert Angert, Bernhard Franzke, Rudolf Maier, and Örjan Skeppstedt to initiate the project CRYRING@ESR is acknowledged.  
The authors are indebted to the ESR team, Markus Steck, Sergey Litvinov, Bernd Lorentz and their colleagues, for providing us with an excellent beam.
We wish to thank also Sonja Bernitt, Chong-Yang Chen, and Ming-Feng Gu for fruitful discussions of the FAC code. This research has been conducted in the framework of the SPARC collaboration, experiment E138 of FAIR Phase-0 supported by GSI. 
It is further supported by the Extreme Matter Institute EMMI and by the European Research Council (ERC) under the European Union’s Horizon
2020 research as well as by the innovation program (Grant No. 682841 “ASTRUm”) and the grant agreement n° 6544002, ENSAR2.
B. Zhu acknowledges CSC Doctoral Fellowship 2018.9 - 2022.2 (Grant No. 201806180051). We acknowledge the support provided by ErUM FSP T05 - "Aufbau von APPA bei FAIR"  (BMBF n° 05P19SJFAA and n° 05P19RGFA1) too.
\end{acknowledgements}
\maketitle

\bibliography{main}

\end{document}